\documentstyle[prl,aps,floats,epsf,color]{revtex}
\begin{document}
\baselineskip=12pt
\def\be{\begin{equation}}
\def\ee{\end{equation}}
\def\bea{\begin{eqnarray}}
\def\eea{\end{eqnarray}}
\def\E{{\rm e}}
\def\bearst{\begin{eqnarray*}}
\def\eearst{\end{eqnarray*}}
\def\peleven{\parbox{11cm}}
\def\peffec{\peight{\bearst\eearst}\hfill\peleven}
\def\pspace{\peight{\bearst\eearst}\hfill}
\def\ptwelve{\parbox{12cm}}
\def\peight{\parbox{8mm}}
\twocolumn[\hsize\textwidth\columnwidth\hsize\csname@twocolumnfalse\endcsname
\title
{Multifractal Detrended Cross-Correlation Analysis of Sunspot
Numbers and River Flow Fluctuations}

\author{S. Hajian, M. Sadegh Movahed}
\address{Department of Physics, Shahid Beheshti University, G.C., Evin, Tehran 19839, Iran}

\vskip 1cm

 \maketitle



\begin{abstract}
We use the Detrended Cross-Correlation Analysis (DCCA) to investigate the influence of
sun activity represented by sunspot numbers on one of the climate indicators, specifically
rivers, represented by river flow fluctuation for Daugava, Holston, Nolichucky and French
Broad rivers. The Multifractal Detrended Cross-Correlation Analysis (MF-DXA) shows that
there exist some crossovers in the cross-correlation fluctuation function versus time scale
of the river flow and sunspot series. One of these crossovers corresponds to the well-known
cycle of solar activity demonstrating a universal property of the mentioned rivers. The
scaling exponent given by DCCA for original series at intermediate time scale, $(12-24)\leq
s\leq 130$ months, is $\lambda = 1.17\pm0.04$ which is almost similar for all underlying rivers at $1\sigma$confidence interval showing the second universal behavior of river runoffs. To remove
the sinusoidal trends embedded in data sets, we apply the Singular Value Decomposition
(SVD) method. Our results show that there exists a long-range cross-correlation between
the sunspot numbers and the underlying streamflow records. The magnitude of the scaling
exponent and the corresponding cross-correlation exponent are $\lambda\in (0.76,
0.85)$ and $\gamma_{\times}\in(0.30, 0.48)$, respectively. Different values for scaling and cross-correlation exponents may
be related to local and external factors such as topography, drainage network morphology,
human activity and so on. Multifractal cross-correlation analysis demonstrates that all
underlying fluctuations have almost weak multifractal nature which is also a universal
property for data series. In addition the empirical relation between scaling exponent
derived by DCCA and Detrended Fluctuation Analysis (DFA), $
\lambda\approx(h_{\rm sun} + h_{\rm river})/2$ is confirmed.

 \end{abstract}
\hspace{.3in}
\newpage
 ]
\section{Introduction}

Recently, due to the developments in the area of complex systems as well as data measurements and data analysis, one can find many opportunities for examination and interpretation of climate change which exhibit irregular systems
\cite{1,1-1,1-2,whit1,whit1-1,whit1-2,mar,sun11}. It is well shown that the climate system is enforced by the well-defined seasonal periodicity, however the existence of
unpredictable perturbation and chaotic functioning lead to extreme climate events. Indeed the climate is a dynamical
system affected by tremendous factors and variables, such as solar activity which is represented by Sunspot numbers in
this research \cite{whit1,whit1-1,whit1-2,mar,sun11,pablo,pablo10}. All factors that control the trajectory of such mentioned systems have enormously large phase space
and evolve as non-stationary processes, consequently we should explore it with stochastic tools to achieve reliable results.
Nowadays, it has been clarified that a remarkably wide variety of natural systems can be characterized by long-rangeSuch cross-correlations address scientists toward fractal geometry of the
underlying dynamical systems and can hopefully help us to predict future events. Existence and determination of power-law cross-correlations would help to promote our understanding of the corresponding dynamics and their future evolutions \cite{val03,livina03,sadeghriver,ser09}. Beside, many events which controls earths climate, water runoff records assigned by rivers and sun activity play a
crucial and survival roles for human life. The runoff water fluctuations are excellent climate criteria because they integrate
evapotranspiration (output) and precipitations (input) over large areas. It is well accepted that the prediction of water runoff
is fundamental for different aspects of social and economical reasons, ranging from the prediction of floods and droughts
to planning of agricultural conditions. As a result of the periodicity in precipitation, river flow has also strong seasonal
periodicity \cite{livina03}. It is worth to note that unlike other climate components, water runoff may be directly influenced by human
activity, like agriculture, drainage network morphology and so on, consequently makes it hard to distinguish the artificial
and natural effects on the river flow data. Finding some or at least a universal behavior for different streamflow fluctuations
as well as quantifying the impact of sun activity on various temporal and spatial scales of water runoff fluctuations can
improve the recent hydrological models \cite{zhang09}.

To this end, the statistical and multifractal analysis of river
flows as well as influence of sun activity due to the interior and
exterior chemical and physical properties of sun should be an
important issue in the geophysical and hydrological systems.

The streamflow of rivers and sun activity have been studied from
various point of views such as: the probability distribution
\cite{mur,kroll}, correlation and fractal behaviors
\cite{Hurst51,Hurst51-1,Hurst51-2,Hurst51-3,Hurst51-4,Hurst51-5},
connection between volatility and nonlinearity of fluctuations
\cite{val03,livina03,Ashkenazy01,Ashkenazy03}, scale invariance for
distribution function \cite{loboda}. In addition, sun activity have
been investigated by some methods in chaos theory \cite{veronig02}
and also multifractal analysis \cite{valentyana05,zhukov03}, wavelet
analysis \cite{ter002}, cross-correlation functions between monthly
mean sunspot areas and sunspot numbers \cite{tem02,bog82}, the
relation between sunspot numbers fluctuation and number of flares,
their evolution step \cite{ter02,ter03}, principal components and
neural network methods to predict sunspots \cite{pc,li}, sunspot
areas time series and solar irradiance reconstructions \cite{balm},
magnetic and dynamic properties of sunspots at the photospheric
level \cite{A. Tritschler} and the hydraulic-geometric similarity of
river \cite{Schmitt,bor,bor-1}.

More recently, Pablo J. D. Mauas et. al., have investigated the
solar forcing on climate, using the quantification of
cross-correlation between the yearly sunspot numbers, irradiance
reconstruction and streamflow of Paran\'{a} river
\cite{pablo,pablo10}.
On the other hand, the mechanisms for solar
influence on the earth's climate has been clarified in detailed from
various point of views in \cite{joa}. Q. Zhang et. al., have
investigated the universal behavior of streamflow records of the
Pearl river \cite{zhang09}.

After innovation of Hurst to propose the self-similar processes and
its criteria, namely ``Hurst exponent"
\cite{Hurst51,Hurst51-1,Hurst51-2,Hurst51-3,Hurst51-4,Hurst51-5},
long-range correlated fluctuation behavior has also been reported
for vast category of sciences, specifically the geophysical records
(for more discussion see
\cite{Hurst51,Hurst51-1,Hurst51-2,Hurst51-3,Hurst51-4,Hurst51-5,Mand1,Mand1-1,Mand1-2,Mand1-3}).
In the last decade, the modification prescription which is required
for a full characterization of many data sets such as the runoff
records, the various moments of the so-called fluctuation functions,
have been introduced
\cite{Hurst51,Hurst51-1,Hurst51-2,Hurst51-3,Hurst51-4,Hurst51-5}.
The effect of non-stationarity on the detrended fluctuation analysis
has been investigated in \cite{kunhu,trend2}.

\begin{figure}[t]
\epsfxsize=9truecm\epsfbox{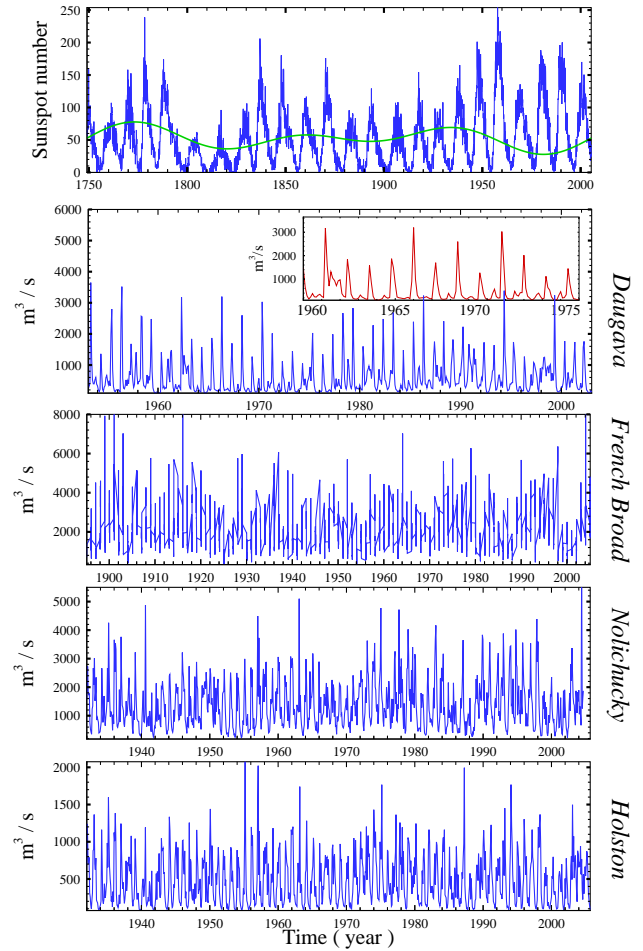} \narrowtext \caption{Upper
panel corresponds to the monthly sunspot number data set. The secular
trend, obtained with a low-pass Fourier filter is shown as a thick
line in the upper panel. Lower panels indicate observed flux
fluctuations of Daugava, French Broad, Nolichucky and Holston
rivers, respectively. The inset plot shows river flow for small scales.}
\label{fig1}
 \end{figure}

Here we take a new approach and rely on the state-of-the-art
algorithm to investigate the contribution of sinusoidal trends
embedded in the data set as well as non-stationarity properties of
the underlying series.
 We implement robust methods to explore the
multifractal nature of cross-correlation between two important
climate variables, the monthly streamflow of some rivers and sun
activity represented by sunspot numbers (see Figure (\ref{fig1})),
by using the novel approach in the fractal analysis, Detrended
Cross-Correlation Analysis (DCCA) and its multifractal modification,
the Multifractal detrended Cross-Correlation Analysis (MF-DXA)
\cite{DCCA,mf-dxa}. We restrict this article to use the sunspot
numbers as the solar activity indicator, since there are many large
and reliable data sets which can be considered as solar influence on
the climate. Due to the presence of the sinusoidal trends in both
sunspot numbers and the runoff river fluctuations and based on
previous researches, one cannot expect to find a unique scaling
behavior for fluctuation functions in all time scales (see section
III), consequently, we have been motivated to use the well-known
method, namely Singular Value Decomposition (SVD) method to exclude
dominant trends in data set (see section II for more details). So after,
clean data set will be used in the DCCA and MF-DXA methods.

This paper is organized as follows: in section II, we describe the
methods which are used to determine the cross-correlation of two
non-stationary time series, the Detrended Cross-Correlation Analysis
(DCCA), and investigate the corresponding multifractal properties by
us- ing the Multifractal Detrended Cross-Correlation Analysis (MF-DXA). Section II will be continued by introducing a method to
eliminate trends from the original data set, the Singular Value
Decomposition (SVD), and describing data used in this paper. In
section III, the multifractal cross-correlation of the underlying
data sets will be examined. Section IV, will be devoted to the results and summary.\\

\section{Analysis techniques AND data description}

Time series measured in the nature are usually affected by
non-stationarities such as trends and artificial noises which must
be well distinguished from the intrinsic fluctuations of the series.
In many cases also, intrinsic fluctuations behave as non-stationary
processes. Consequently, common methods in data analysis will be
encountered with spurious or at least unreliable results. One of the
most famous and well-known approach used in many studies is
Multifractal Detrended Fluctuations Analysis (MF-DFA)
\cite{Peng95,bun02}. This method has been applied to various areas,
such as 
 economical
time series \cite{economics,economics-1,economics-2,economics-3,economics-4}, river flow \cite{sadeghriver} and
sunspot fluctuations \cite{sadeghsun,husun}, cosmic microwave background
radiations \cite{sadeghcmb}, music \cite{jafarimusic1,jafarimusic1-1}, plasma
fluctuations \cite{movahedplasma}.

For many reasons, we are interested in studying the mutual influence
of two series in the presence of non-stationarities. Obviously,
traditional methods for this investigation become inaccurate
procedures. Recently Jun et. al. have proposed an approach for
analyzing correlation properties of a series by decomposing the
original signal into its positive and negative fluctuation
components \cite{woo}. Based on the previous study, Podobnik et. al.
have modified the mentioned correlation method and improved it to
explore the cross-correlation between two non-stationary
fluctuations, named Detrended Cross-Correlation Analysis (DCCA)
\cite{DCCA} and its generalized, the Multifractal Detrended
Cross-Correlation Analysis (MF-DXA) which also examine higher orders
detrended covariance\cite{mf-dxa}.

As mentioned before, trends in data set may influence the accuracy
of results. For reliable detection of the cross-correlations, it is
essential to distinguish trends from the intrinsic fluctuations in
data. Generally, trends embedded in measurements are of two types:
Polynomial and Sinusoidal trends. Although the MF-DFA and MF-DXA
methods eliminate the polynomial trends, the sinusoidal trends
remain \cite{kunhu,trend2}. There are several robust methods to
eliminate the sinusoidal one such as Fourier Detrended Fluctuations
Analysis (F-DFA) \cite{movahedplasma,f-dfa}, which is actually a
high-pass filter, and Singular Value Decomposition (SVD)
\cite{trend3,trend3-1}. One of the most disadvantage of the F-DFA
method is the reduction of the size of underlying data set.
To apply
the MF-DXA (see the following subsection), we have to synchronize
two underlying data set which may not be done using the F-DFA
method. On the other hand, the SVD method promises to remain the
length as well as synchronization \cite{trend3,trend3-1}. Here in
order to eliminate the effect of sinusoidal trends, we apply the
Singular Value Decomposition (SVD) \cite{trend3,trend3-1}. After
trends elimination, we use the MF-DXA to analyze the cleaned data
sets.

\subsection{DCCA and MF-DXA}
One of the newly methods in analyzing two non-stationary time series
is Detrended Cross-Correlation Analysis (DCCA)\cite{DCCA,woo}. This
method is a generalization of the DFA method in which only one time
series was analyzed. Recently a generalized version of the DCCA
method which is so-called MF-DXA, has been introduced
\cite{mf-dxa}. Just same as the MF-DFA method, MF-DXA consists of
the $4$ steps (see \cite{DCCA,mf-dxa,rezadcca09}
for more details):\\
(I): Computing the profiles of the underlying data series, $x_k$ and
$y_k$, as
\begin{eqnarray}
X(i) &\equiv& \sum_{k=1}^i \left[ x_k - \langle x \rangle \right]
\qquad i=1,\ldots,N \nonumber\\
Y(i) &\equiv& \sum_{k=1}^i \left[ y_k - \langle y \rangle \right]
\qquad i=1,\ldots,N \label{profile}
\end{eqnarray}
the subtraction of mean is not compulsory. Since we are going to
compare two different time series, we construct data sets with zero
mean and unit variance using initial ones.

(II): Dividing each of the profile into $N_s \equiv {\rm int}(N/s)$
non-overlapping segments of equal lengths $s$, and then computing
the fluctuation function for each segments. In order to take the
whole series into account when the size of the data sets is  not a
multiple of considered time scale, $s$, we do the same procedure
from the opposite end, consequently one finds $2N_s$ segments.
\begin{eqnarray}
&&F(s,m) ={1 \over s} \sum_{i=1}^{s}\{Y[(m-1) s+ i] - y_{m}(i)\}
\nonumber\\&&\qquad\qquad\qquad\qquad\times\{X[(m-1) s + i] -
x_{m}(i)\} \label{fsdef1}
\end{eqnarray}
for $m=1,...,N_s$ and:
\begin{eqnarray}
&&F(s,m) ={1 \over s} \sum_{i=1}^{s}\{Y[N-(m-1) s+ i] - y_{m}(i)\}
\nonumber\\&&\qquad\qquad\qquad\qquad\times\{X[N-(m-1) s + i] -
x_{m}(i)\} \label{fsdef2}
\end{eqnarray}
for $m=N_s+1,...,2N_s$, where $x_{m}(i)$ and $y_{m}(i)$ are a
fitting polynomial in segment $m$th. Usually, a linear function is
selected for fitting the function. If there is no trend in the data,
a zeroth-order fitting function
might be enough \cite{PRL00}.\\

 (III): Averaging the local fluctuation function over all the part, given
by:
\begin{equation}
F_q(s) =\left\{ {1 \over  N_s} \sum_{m=1}^{ N_s} \left[ F(s,m)
\right]^{q/2} \right\}^{1/q} \label{fdef}
\end{equation}
Generally, $q$ can take any real value, except zero. For $q=0$,
equation (\ref{fdef}) becomes:
\begin{equation}
F_0(s)= \exp\left( {1 \over  2N_s} \sum_{m=1}^{ N_s}\ln F(s,m)\right
) \label{fdef0}
\end{equation}
For $q=2$, the standard DCCA is retrieved.\\

(IV): The final step is  calculating the slope of the log-log plot
of $F_q(s)$ versus $s$ which directly determines the scaling
exponent $\lambda(q)$, as:
\begin{equation}
F_q(s) \sim s^{\lambda(q)} \label{Hq}
\end{equation}
If both underlying series are equal then $\lambda(q)$ is nothing
else but so-called generalized Hurst exponent, $h(q)$.  In the
absence of sinusoidal trends embedded in data sets, if one finds no
scaling behavior for the fluctuation function in equation (\ref{Hq})
or at least, there does not exist any unique exponent for all
scaling ranges then there exists either short-range cross-correlation or not at all any cross-correlation. For a series of
size $N$,  the minimum number of windows will be $N_s=2$
corresponding to the maximum value of $s={\rm int}(N/2)$. In
addition, it has been demonstrated that to find the most correct
value of scaling exponent by using DFA and DCCA methods, we should
set $s\leq (N/2)$, namely $N_s\ge 2$ \cite{bun02}.

To determine the slope of curve in the log-log plot of fluctuation
function versus scale (equation (\ref{Hq})), we use Bayesian
statistics\cite{fab04} . We introduce measurements and model
parameters as $\{X\}:\{F_q(s)\}$ and $\{\Theta\} : \{\lambda(q)\}$,
respectively. Based on the Bayesian theorem, the conditional
probability of the model parameters given data set (observation) is
so-called posterior probability and is given by:

\begin{equation}
P(\lambda(q)|X)=\frac{{\mathcal{L}}(X|\lambda(q))P(\lambda(q))}{\int
{\mathcal{L}}(X|\lambda(q))d\lambda(q)}
\end{equation}
where the first term in the nominator of the right hand side is
Likelihood and the second term contains all initial constraints
concerning model parameters, so-called prior distribution. This term
expresses the degree of belief about the model. In the absence of
every prior constraints, the posterior function, $P (\lambda(q)|X )$
is proportional to the Likelihood function. If there is no
correlation between various measurements, consequently according to
the central limit theorem, Likelihood function is given by a product
of Gaussian functions as follows:

\begin{equation}
{\mathcal{L}}(X|\lambda(q))\sim\exp\left(\frac{-\chi^2(\lambda(q))}{2}\right)
\end{equation}
 where:
\begin{equation}\chi^2(\lambda(q))=\int ds\frac{[F_{{\rm obs.}}(s)-F_{{\rm The.}}(s;\lambda(q))]^2}{\sigma_{{\rm obs.}}^2(s)}\end{equation}
Here $F_{{\rm obs.}}(s)$ and $F_{{\rm The.}}(s;\lambda)$ are
fluctuation function computed directly from the data set by using
DFA or DCCA and determined by equation (6), respectively. Also,
$\sigma_{{\rm obs.}}(s)$ is the mean standard deviation, associated
to $F_{\rm obs.}(s)$. Apparently, this Likelihood function to be
maximum when for a value of the scaling exponent, $\lambda(q)$,
$\chi^2$ reaches to its global minimum. The value of error-bar at
$1\sigma$ confidence interval of $\lambda(q)$ is determined by the
likelihood function based on the following condition:
\begin{equation}
68.3\%=\int_{-\sigma^{-}}^{+\sigma^{+}}{\mathcal{L}}(X|\lambda(q))d\lambda(q)
\end{equation}

Finally we report the best value of scaling exponent at $1\sigma$
confidence interval according to $\lambda_{-\sigma^-}^{+\sigma^+}$
for each moment, $q$'s.

\subsection{Singular Value Decomposition (SVD)}

Determining trends and construction proper detrending operations are
important step toward robust analysis, specially in climatic data
analysis. As given by Z. Wu and his collaborators \cite{wu07}, there
is no unique definition of trend and any proper algorithm for
extracting it from underlying  stationary as well as non-stationary
data sets. In another aspect, the trend in a real world data series,
non-stationary one, is an intrinsic function imposed by the nature
on data set. To identify the trend on a data set, we can investigate
the series in whole domain or on some specific span of domains. For
linear and stationary data sets choosing the length of data set as
domain of trend may be suit but for a real world data set which is
non-stationary and nonlinear, we need more precise definition of
trend.
 As of the importance of investigating trends and probably
removing them from series, one can point out to two following aims:
\\ I) In some cases, there exists one or more crossover (time)
scales, $s_\times$, in the log-log plot of $F_q(s)$ versus $s$
(equation (\ref{Hq})), segregating regimes with different scaling
exponents. These patterns demonstrates that underlying fluctuation
has different correlation behavior in
various values of scales \cite{kunhu,trend2,trend3,trend3-1,physa}.   \\
II) In many cases, crossovers are produced by the embedded
sinusoidal trends, e.g. seasonal trends in the climate time series.
Subsequently, to find scaling exponent of the intrinsic
fluctuations, we should remove sinusoidal trends by using most
robust detrending method so after, produced clean data set is used
as an input for DCCA and MF-DXA methods
\cite{kunhu,trend2,movahedplasma}.

In order to remove trends corresponding to the low frequency
periodic behavior, we use Singular Value Decomposition method
\cite{golub} instead of transformation a recorded data to the
Fourier space using the method proposed in \cite{cooly65} (see also
\cite{kunhu,trend2,physa}). Using the SVD method not only we can
track the influence of sinusoidal trends on the results, but also
the synchronization of two underlying data sets will be guaranteed
\cite{trend3,trend3-1}. In addition, we can determine over which
scale, noises or trends have dominant contribution also the value of
so-called crossover in the fluctuation function, in terms of the
scale is computed by using DCCA method.
\cite{f-dfa,trend3,trend3-1,SVD,koscielny98}. After removing the
dominant periodic functions, such as sinusoidal trends, we obtain
the fluctuation exponent by direct application of the
MF-DXA.\\
The SVD method consists of the following steps:

(I): Consider a data set which is superimposed with periodic trends
$\{x_i\}; i=1, ..., N$. Embed $x_i$ with parameters $(d,\tau)$ where
$d$ and $\tau$ are the embedding dimension and the time delay,
respectively. The embedded data can be represented as a matrix
$\mathbf{\Gamma}$ given by:
\begin{equation}
\mathbf{\Gamma}\equiv\left(
  \begin{array}{cccc}
    x_1 & x_{1+\tau} & ... & x_{1+N-(d-1)\tau-1} \\
   \vdots&\vdots&\vdots &\vdots \\
      x_i & x_{i+\tau} & ... & x_{i+N-(d-1)\tau-1} \\
   \vdots&\vdots&\vdots &\vdots \\
  x_d & x_{d+\tau} & ... & x_{d+N-(d-1)\tau-1} \\
  \end{array}
\right)\label{matrix1}
 \end{equation}
as one can see, $1 \leq i\leq d$.
For a periodic embedded trend, we
set $p$ as the number of dominant frequency components of the form
$e^{({\rm i} \omega_k t)}$, $k = 1 ... p$ in their power spectrum.
For a fixed value of sample size, $N$, the maximum value of  $d$ in
the so-called trajectory matrix, $\mathbf{\Gamma}$, equates to
$d\leq N-(d-1)\tau+1$ \cite{trend3-1,SVD,shang09}.

\begin{figure}[t]
\epsfxsize=9cm\epsfbox{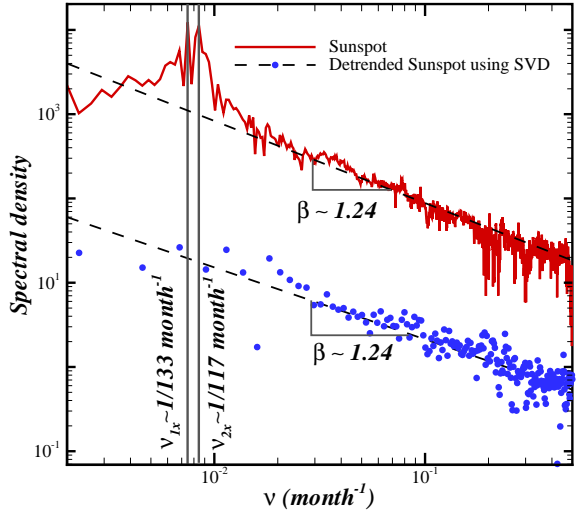} \caption{Power spectrum of the original
sunspot numbers data set (solid line) and that of for a cleaned one
using SVD method (filled symbols). Dashed lines correspond to the
scaling function as $\nu^{-\beta}$ for the same signals without
sinusoidal trends.} \label{power}
 \end{figure}

(II): Matrix $\mathbf{\Gamma}$ will be decomposed to the two left
and right orthogonal matrices as follows:
\begin{equation}
\mathbf{\Gamma}={\mathbf{USV^{\dagger}}}
\end{equation}
where ${\mathbf{U}}_{d\times d}$ and
${\mathbf{V}}_{(N-(d-1)\tau)\times(N-(d-1)\tau)}$ are the left and
right orthogonal matrices, respectively. The diagonal elements of
${\mathbf{S}}_{d\times (N-(d-1)\tau)}$ are the desired singular
values, also known as eigenvalues. The Singular Value Decomposition
of $\mathbf{\Gamma}$ is related to the eigendecomposition of the
symmetric matrices $\mathbf{\Gamma}^\dagger \mathbf{\Gamma}$ and $
\mathbf{\Gamma} {\mathbf{\Gamma}}^\dagger$, as
${\mathbf{\Gamma^\dagger \Gamma v_i}}=\lambda_i^2 \mathbf{v_i}$ and
${\mathbf{ \Gamma \Gamma^\dagger u_i}}=\lambda_i^2 \mathbf{u_i}$.
The nonzero eigenvalues of $\mathbf{\Gamma^\dagger \Gamma}$ are the
same as that of $\mathbf{\Gamma \Gamma^\dagger }$; and determine the
rank of $\mathbf{\Gamma}$. The eigenvalues are ordered such that $
\lambda_i> \lambda_{i+1} \geq 0$. The diagonal elements of
$\mathbf{S}$ will be constructed by the ordered eigenvalues (the
others will be set to zero). The rank of $\mathbf{\Gamma}$ is equal
to the number of nonzero eigenvalues. The columns of $\mathbf{U}$
and $\mathbf{V}$ are constructed by the $\mathbf{u_i}$ and
$\mathbf{v_i}$, namely the eigenvectors mentioned before,
corresponding to the ordered eigenvalues.

By applying the SVD to the matrix $\mathbf{\Gamma}$ (i.e.,
$\mathbf{\Gamma}=U S V^\dagger$) we will get $\mathbf{U}$,
$\mathbf{S}$ and $\mathbf{V}$. We consider the number of frequency
components in the periodic trend to be $p$. Set the dominant $2p+1$
eigenvalues in the matrix $\mathbf{S}$ to zero and hereafter named
as $\mathbf{S^*}$. The filtered matrix ${\mathbf{\Gamma}^*}_{d
\times(N-(d-1)\tau)}$ determined by $\mathbf{\Gamma}^* = U S^*
V^\dagger$ with elements $\Gamma^*_{ij}$. This in turn is mapped
back on to a one-dimensional or filtered data given by:

\begin{figure}[t]
\epsfxsize=9cm\epsfbox{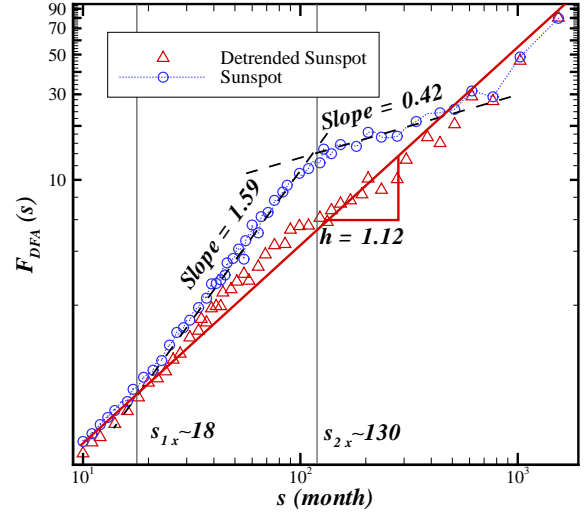} \caption{The DFA results of
the original sunspot number data set and filtered one using SVD. The solid
line corresponds to power fit of fluctuation. The slope of power
fits has been calculated by likelihood analysis (see the text).}
\label{sun}
 \end{figure}

\begin{equation}
x^*_{i+j-1}=\Gamma^*_{ij}
\end{equation}
where $1\leq i \leq d$ and $ 1\leq j \leq N-(d-1)\tau$. The $p$
dominant eigen-values and associating eigendecomposed vectors,
represent trends subspace subsequently, the remaining $(d-p)$
eigenvalues and corresponding eigenvectors demonstrate intrinsic
fluctuations subspace. In order to determine the value of $p$ for a
typical series such as monthly sunspot data set, at first, we
compute the power spectrum of monthly sunspot numbers. As shown in
Figure (\ref{power}), there are at least two dominant sinusoidal
trends embedded in the monthly sunspot numbers. The first one
corresponds to the well-known sun activity and second period belongs
to the so-called Schwabe cycle interval. In order to eliminate
mentioned trends, we set $p=2$ in the SVD algorithm and compute the
power spectrum of the cleaned data again. Our expectation is that
there must be no deviation from scaling function as $\nu^{-\beta}$
with $\beta=2h(q=2)-1=1.24\pm0.02$ as one can see in the lower plot
in Figure (\ref{power}). It is worth to note that by increasing $p$
from its optimum value in the SVD method, probably some intrinsic
statistical properties of underlying data set will be destroyed.
After which we use the detrended sunspot series as an input data for
common DFA method (see Figure (\ref{sun})). As shown in Figure
(\ref{sun}), all of the crossover time scales which are produced due
to the competition between sinusoidal trends embedded in sunspot
series and intrinsic fluctuations, were diminished and intrinsic
fluctuations will be retrieved. Our result is in agreement with the
previous result regarding to scaling behavior of sunspot based on
MF-DFA method accompanying by Fourier-Detrended Analysis, reported
in \cite{sadeghsun}. Figure (\ref{sun}) confirms that, SVD method
could remove sinusoidal trends.

\begin{figure}[t]
\epsfxsize=9cm\epsfbox{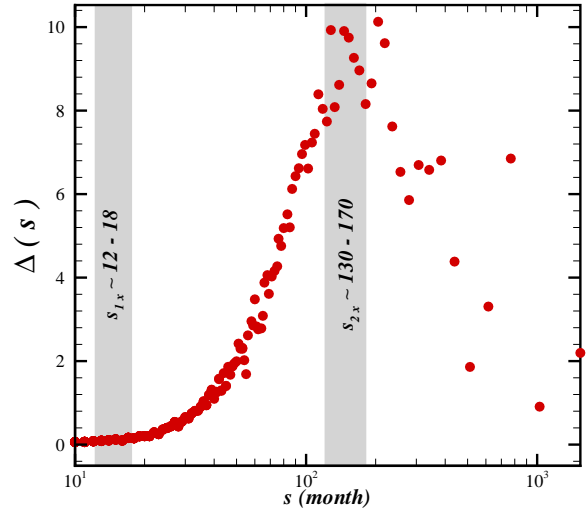} \narrowtext \caption{$\Delta$
function versus time scale. }
 \label{crossover}
 \end{figure}

In the log-log plot of fluctuation function versus time scale given
by DFA method, also one can find three crossovers (see Figure
(\ref{sun})). To determine their value, we define error function as:
\begin{eqnarray}
\Delta(s)=\sqrt{\left[F_{\rm obs.}(s)-F_{\rm{Linear}}(s)\right]^2}
\label{chi_cross}
\end{eqnarray}
for each $q$, where $F_{\rm obs.}(s)$ and $F_{\rm{Linear}}(s)$ are
the fluctuation functions for the original data and the filtered
data produced by the SVD method, respectively. In Figure
(\ref{crossover}) we plot $\Delta$ as a function of $s$ for the
sunspot numbers fluctuations. The first crossovers occurs at
$s_{1\times}\sim [12-18]$ months corresponding to the annual period.
The second crossovers is equal to $s_{2\times}\sim [130-170]$ months
which is related to the well-known solar activity period. We cannot
determine the value of third crossover with the acceptable
uncertainty by using DFA method due to small size of current sunspot
series.

\begin{figure} [t]
\epsfxsize=9cm\epsfbox{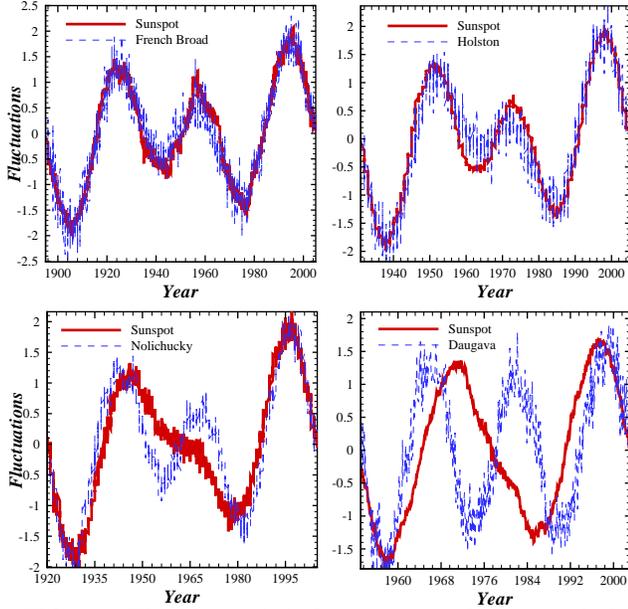} \narrowtext \caption{The detrended
fluctuations of rivers (dashed line) and the corresponding sunspot
(solid line) time series. Here we use a SVD method to trend
series. The mean and variance of
each series became zero and unity, respectively. The Pearson's
correlation coefficients are $r_{\rm Daugava}=+0.44$, $r_{\rm Holston}=+0.94$, $r_{\rm Nolichucky}=+0.88$ and $r_{\rm French Broad}=+0.94$.}
 \label{fluc}
 \end{figure}

\begin{figure} [t]
\epsfxsize=9cm\epsfbox{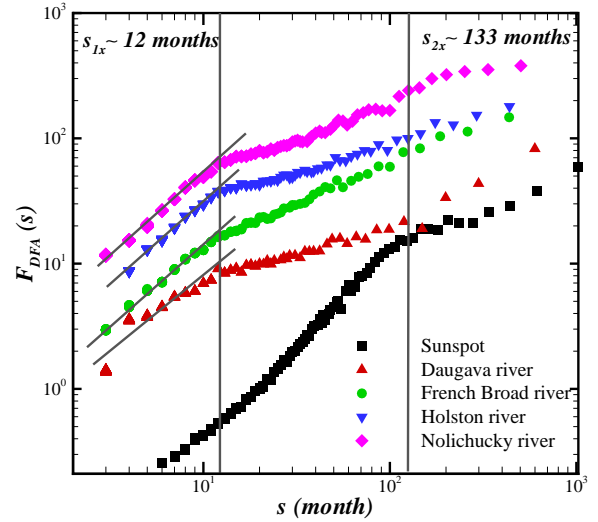} \narrowtext \caption{Fluctuation
function versus scale in log-log plot computed by DFA method for
original data sets. To make more sense we shifted the vertical
scale.}
 \label{dfa}
 \end{figure}

\begin{figure}[t]
\epsfxsize=9cm\epsfbox{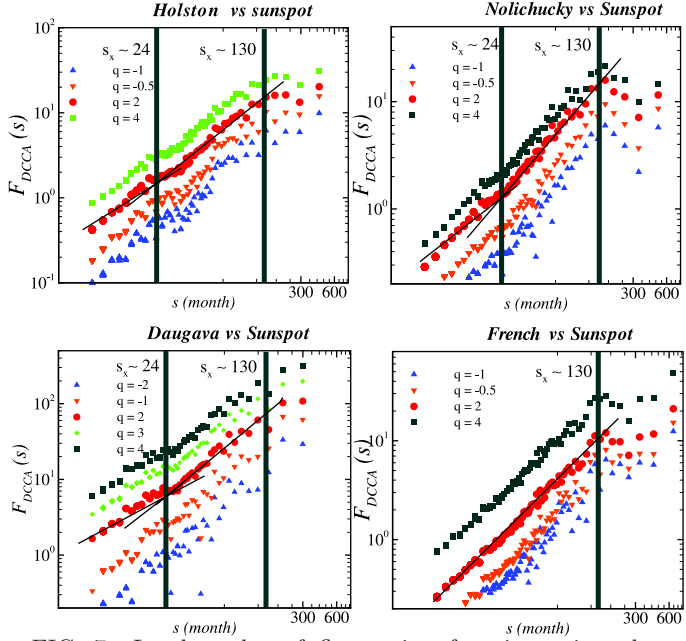} \narrowtext \caption{Log-log plot
of fluctuation functions given by the MF-DXA ($F_{DCCA}(s)$) method for
various values of $q$ for four rivers' flow mentioned in the
manuscript versus sunspot numbers as a function of time scale. Here
$s_{max}=[{\rm size}\quad{\rm of}\quad {\rm series}/2]$. We shifted
the vertical scale to make more sense. }
 \label{all_q}
 \end{figure}

\begin{figure}[t]
\epsfxsize=9cm\epsfbox{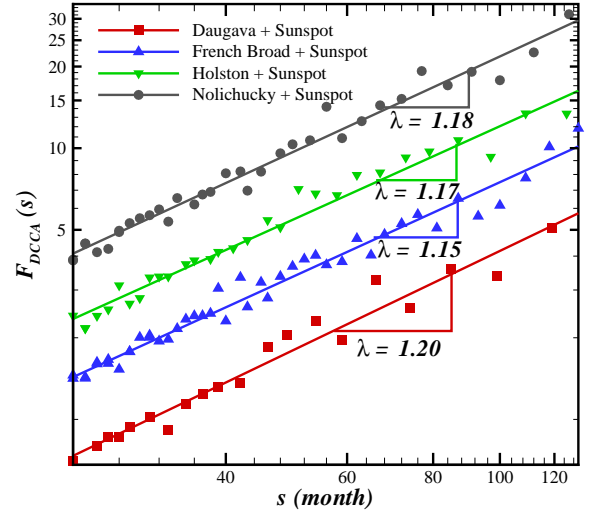} \narrowtext \caption{Log-log plot
of fluctuation functions given by DCCA($F_{DCCA}(s)$) method for
four streamflows fluctuations versus sunspot numbers
as a function $s$ at intermediate time scale, $12-24\leq
s\leq 130$ months. For clearly, we shifted the vertical scale. Solid
lines correspond to the power fit of fluctuation function and slopes
have been computed by likelihood analysis.}
 \label{fs_q}
 \end{figure}

\begin{figure}[t]
\epsfxsize=9cm\epsfbox{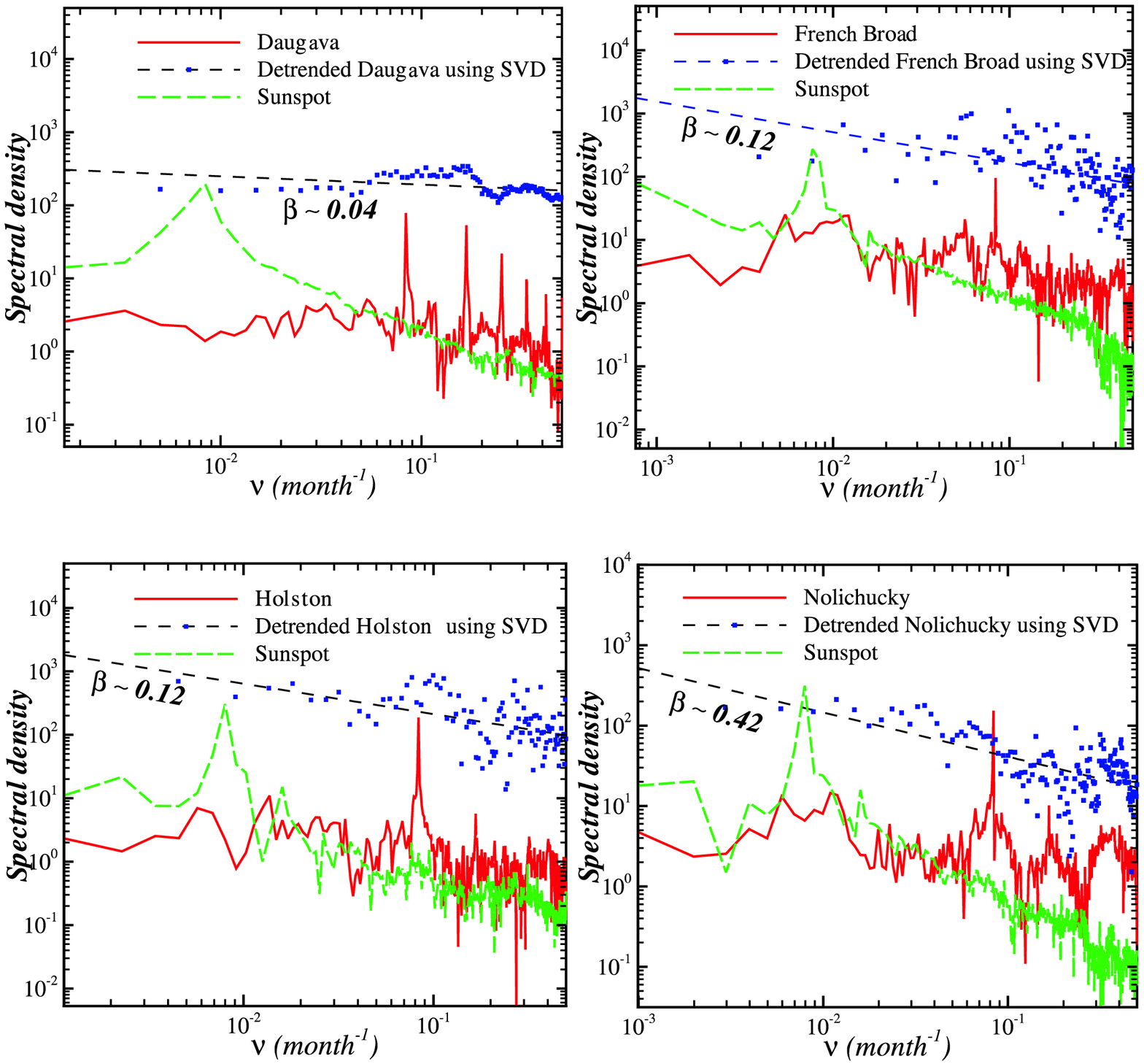} \narrowtext \caption{Spectral
density of river flow fluctuations, detrended data set and
synchronizing sunspot numbers. We shifted the vertical scale for
spectral density to make more sense.}
 \label{power_all}
 \end{figure}

\subsection{Data Description}
We use the up-to-date monthly sunspot numbers ($S_N$) data series
released by National Oceanic and Atmospheric Administration (NOAA)
\cite{1sn} and the Sunspot Index Data Center (SIDC) \cite{data}. The
monthly flow fluctuations of four famous rivers, namely Daugava at
Latvia, Holston near Damascus, Nolichucky at Embreeville and French
Broad at Asheville were collected from National Water Information
System \cite{data_river}. The original Daugava river data source is
Latvian Environmental Geological and Meteorological Agency database
\cite{dau}. The runoff dimension of the underlying rivers is m$^3$/s
(see Figure (\ref{fig1})). The river flow fluctuations have been
measured independently in corresponding area. All mentioned rivers
are mixed feeding from rain, snowmelt water and groundwater. Table
(\ref{Tab_river}) reports some characteristics of river flow
fluctuations used in this study. These data sets are almost long in
length and most available series. In Figure (\ref{fluc}) we plot
underlying streamflows and sunspot numbers detrended data sets. In
this figure, we use SVD method explained in the previous subsection
for detrending data sets, also the mean and variance of series
equate to zero and unity, respectively. According to Figure
(\ref{fluc}), one may find a remarkable correlation for streamflows
and sunspot numbers. To quantify this correlation we use the
Pearson's correlation coefficients for ${X}$ and ${Y}$ defined by:
$r=\frac{\langle X-\langle X\rangle\rangle\langle Y-\langle Y\rangle
\rangle}{\sqrt{\langle X-\langle X\rangle\rangle \langle Y- \langle
Y\rangle\rangle}}$. The value of $r$ for detrended river flows and
sunspot numbers correspond to $r_{\rm Daugava}=+0.44$, $r_{\rm
Holston}=+0.94$, $r_{\rm Nolichucky}=+0.88$ and $r_{\rm
French-Broad}=+0.94$.

\begin{table}[htp]
\caption{\label{Tab_river}The main characteristics of runoff rivers
used in this paper.}
\begin{center}

\begin{tabular}{|c|c|c|c|c|}
    River & Discharge & Series Length& Drainage & Location\\
&(m$^3/$sec)&& area(km$^2$)&\\\hline
  French Broad  &$2093$ &$1896.1-2005.12$&2448&$35^{\circ}36'$ N\\
  &&&& $82^{\circ}34'$ W\\\hline
   Daugava  &$601$&$1953.1-2002.12$&87900& $56^{\circ}57'$ N \\
   &&&&$24^{\circ}6'$ E \\\hline
  Holston  &$479$& $1931.1-2005.12$&784.8&$36^{\circ}39'$ N \\
  &&&&$81^{\circ}50'$ W\\\hline
    Nolichucky &$1379$& $1921.1-2005.12$&2085&$36^{\circ}10'$ N  \\ 
&&&&$82^{\circ}27'$ W
\end{tabular}
\end{center}
\end{table}


\section{MF-DXA of Sunspots and River flow fluctuations}
To examine the multifractal properties and cross-correlation of
sunspot numbers and river flow fluctuations, we apply the DCCA and
MF-DXA methods. As mentioned in the previous section, we have to
synchronize two time series of interests. The length of the river
flow fluctuations may vary from 600 to 1300 months for the studied
rivers, Daugava, Nolichucky, Holston and French Broad. On the
contrary, there is more information available for sunspot databases,
ranging from daily to annually data sets. To find  reliable results
given by DCCA, we synchronized the sunspot monthly series to all
four river's data sets.

Before applying any detrending program we apply DFA, DCCA and MF-DXA
method. Log-log plot of $F(s)$ versus $s$ for original streamflow
and sunspot data sets by using DFA method (Figure (\ref{dfa}))
confirms that all underlying runoff rivers behave as the
anti-persistent long-range correlated series for time scale $s<12$
months. the value of Hurst exponent at this scale for Daugava,
Holston, Nolichucky, French Broad and sunspot correspond to $H_{\rm
Daugava}=h(2)-1= 0.21\pm 0.02$, $H_{\rm Holston}=h(2)-1=
0.22\pm0.02$, $H_{\rm Nolichucky}=h(2)-1=0.21\pm0.02$, $H_{\rm
French- Broad}=h(2)-1= 0.18\pm0.03$ and $H_{\rm
Sunspot}=h(2)-1=0.11\pm0.03$, respectively.

Figure (\ref{all_q}) shows the fluctuation function given by
multifractal DCCA method for each river flow fluctuations versus
sunspot numbers without applying any detrending procedure. Some
crossover time scales in fluctuation functions are detected in
Figure (\ref{all_q}). One of those crossovers is known as cycle of
solar activity. In addition the scaling behavior of fluctuation
function at intermediate time scales, namely $[12-24]\leq s\leq 130$
months, is similar at $1\sigma$ confidence interval for $q=2$ for
all rivers used in this paper (see Figure (\ref{fs_q})). The scaling
exponent at this regime is $\lambda(q=2)=1.17\pm0.04$. As shown in
Ref. \cite{trend2}, trends become dominant at intermediate regimen
while at small and very large time scales, the intrinsic
fluctuations to be dominated, therefor, we conclude that at
$[12-24]\leq s\leq 130$ months the cross-correlation of water runoff
and corresponding sunspot numbers determined by DCCA method for
$q=2$, is characterized as the universal behavior. For $s<[12-24]$
and $s>130$ months the slopes of fluctuation functions are different
for various rivers and implies that at mentioned time scales the
local effects such as morphology, human activities, various drainage
areas become dominant \cite{zhang09,Zanchettin08}. Moreover, cycle
of solar activity represented by sunspot is one of most robust
effect which affects on streamflow fluctuations.

\begin{figure} [t]
\epsfxsize=9cm\epsfbox{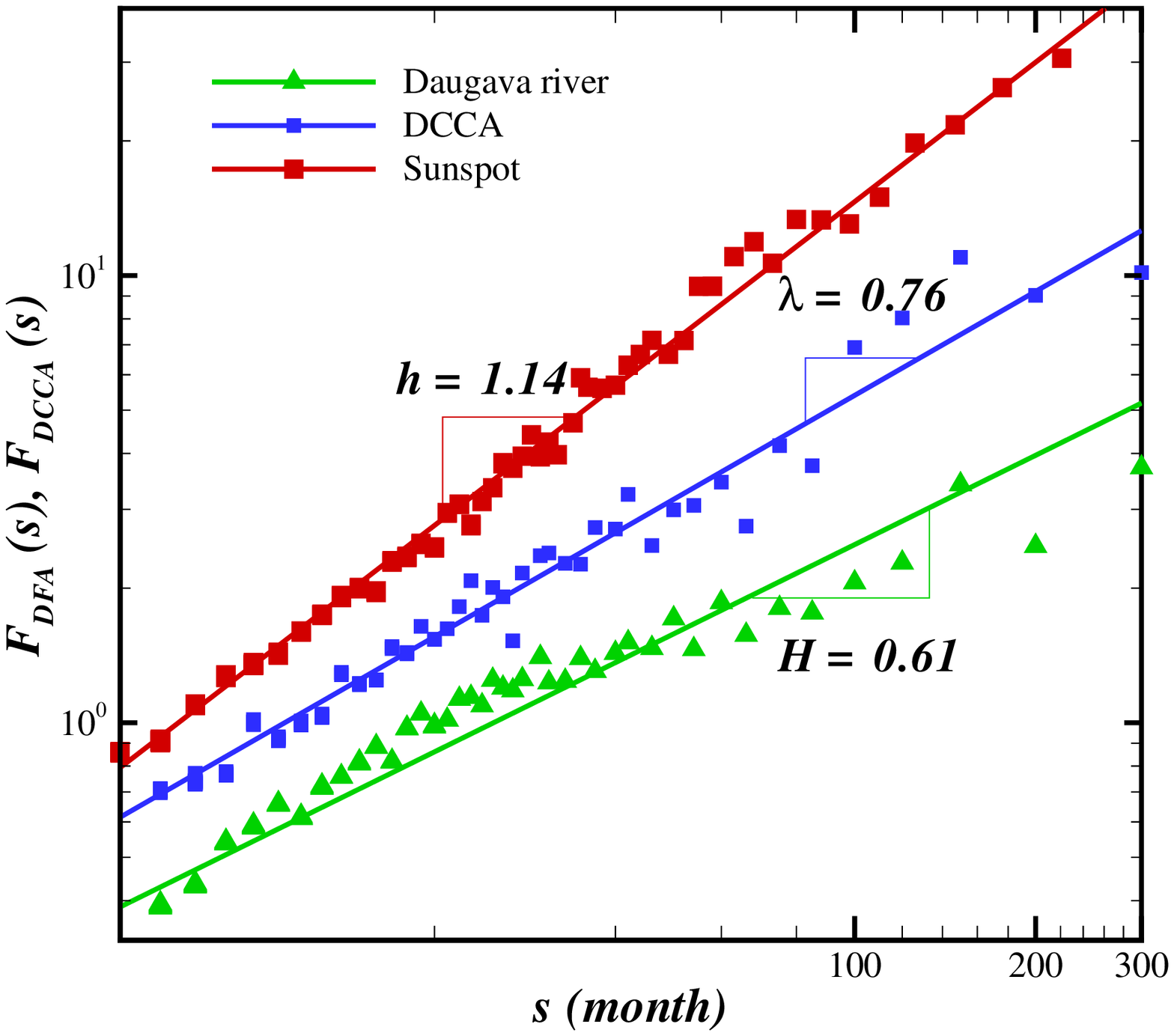} \narrowtext \caption{Log-log plot
of fluctuation functions given by DFA ($F_{DFA}(s)$) and DCCA
($F_{DCCA}(s)$) methods for detrended sunspot numbers and Daugava
river flow as a function of time scale. Where we choose $q=2$. The
slope of power fits have been derived by likelihood statistics.}
 \label{dcca1}
 \end{figure}
\begin{figure} [t]
\epsfxsize=9cm\epsfbox{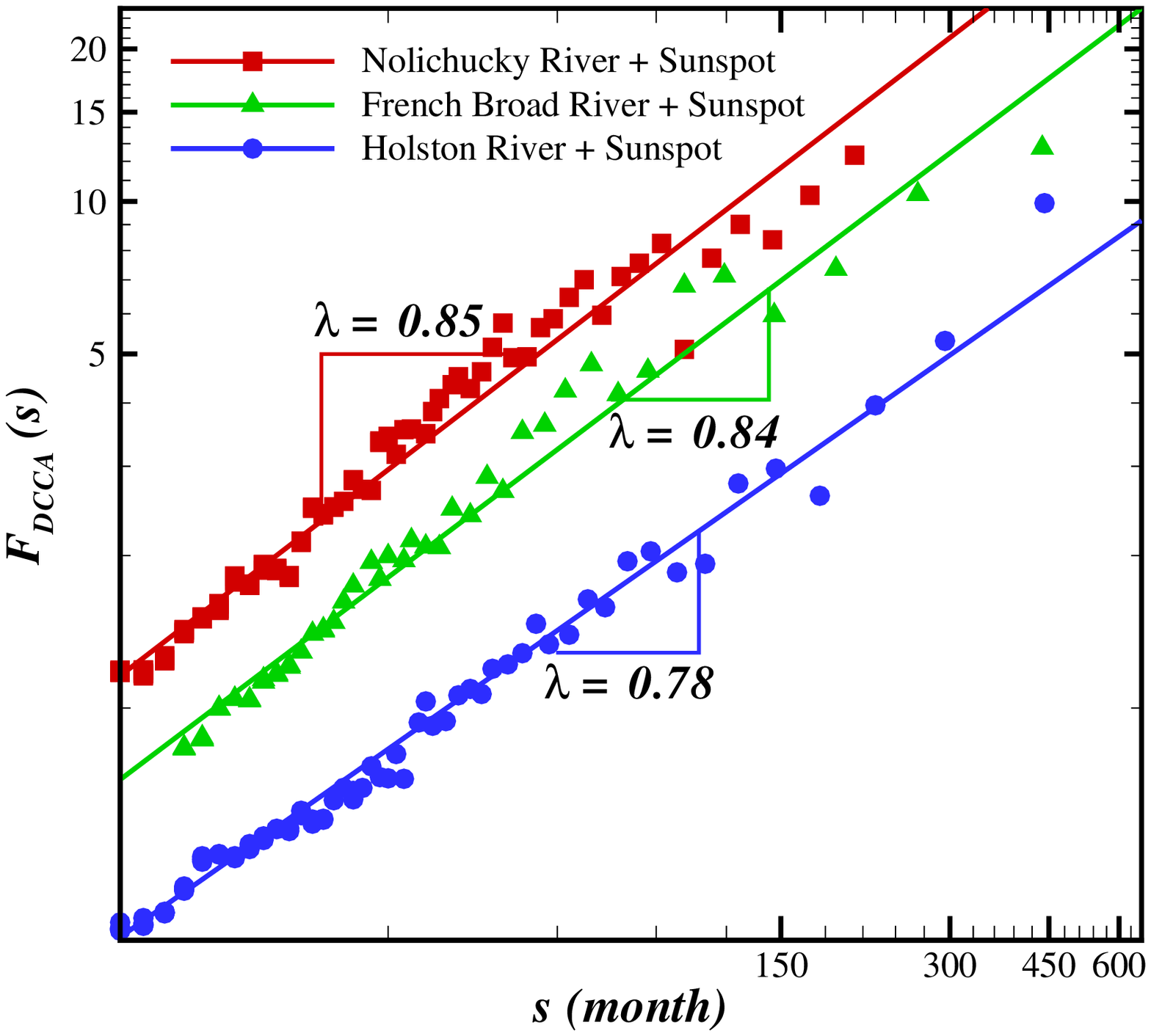} \narrowtext \caption{Fluctuation
function given by DCCA method for detrended Holston, French Broad
and Nolichucky rivers with their synchronized sunspot series in
log-log scale versus time for $q=2$. The maximum value of $s$ for
each series equates to the integer part of  half size of the
corresponding river flow. }
 \label{dcca2}
 \end{figure}

Figure (\ref{power_all}) shows the power spectrum of original and
detrended mentioned rivers data sets and synchronized sunspot numbers
series. The spectral density of French Broad and Nolichucky rivers
behave in similar way around the well-known solar activity. This
outcome is also confirmed in Figure (\ref{all_q}), since there is a
sharp crossover in the fluctuation function versus $s$, in the
MF-DXA method for French Broad and Nolichucky rivers. This may
demonstrate that the contribution of solar activity depends on the
geographical properties of river.

\begin{figure} [t]
\epsfxsize=9cm\epsfbox{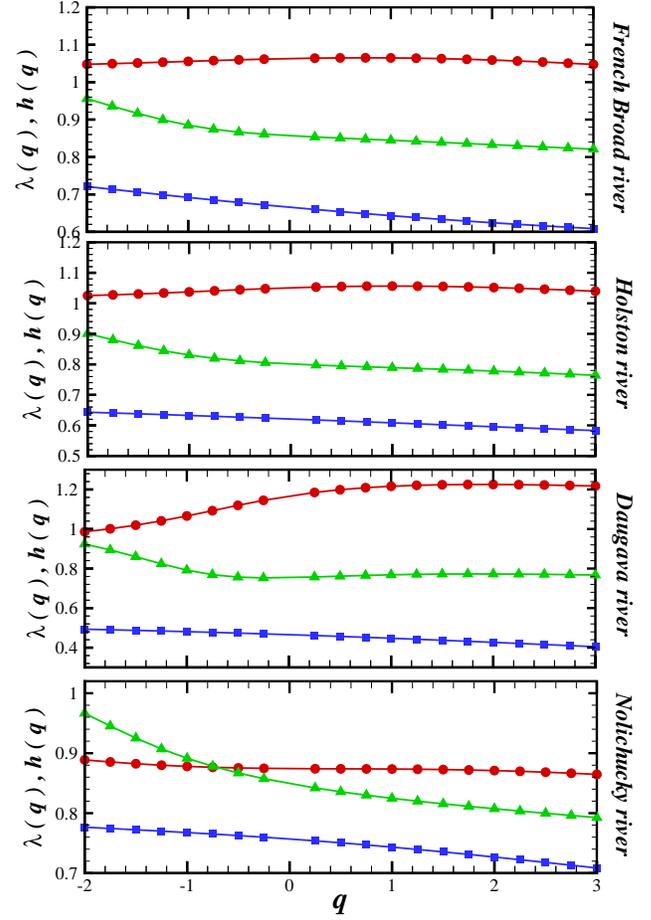} \narrowtext \caption{Generalized
Hurst exponent, $h(q)$ computed by MF-DFA (filled circle and square
symbols) and $\lambda(q)$ derived by equation (\ref{Hq}) in MF-DXA
method (filled triangle symbol). Filled circle and square correspond
to sunspot numbers and mentioned runoff river fluctuations,
respectively.}
 \label{dcca3}
 \end{figure}

Now, to explore the existence of cross-correlation in data set
without sinusoidal trends from the DCCA and MF-DXA method we detrend
each of the above synchronized data set by using the SVD method.
Applying the DCCA method, we find out that there exists an strong
long-range cross-correlation between sun activity represented by
sunspot numbers and river flow fluctuations. Figure (\ref{dcca1})
shows the log-log plot of fluctuation functions introduced by DFA
($F_{DFA}(s)$) for a typical river flow, namely Daugava and sunspot
fluctuations in the same time interval as well as the same function
given by DCCA ($F_{DCCA}(s)$) for those synchronized ones and $q=2$.
Figure (\ref{dcca2}) indicates the results of DCCA method for three
other rivers with their synchronized sunspot series. The scaling
behavior of these functions represented in Figures (\ref{dcca1}) and
(\ref{dcca2}) confirms that there exists cross-correlation between
such river flow fluctuations and sun activity. To explore the
multifractal nature of underlying data sets, we apply MF-DFA
\cite{bun02,sadeghsun} and MF-DXA \cite{mf-dxa} methods. Figure
(\ref{dcca3}) shows generalized Hurst exponent and $\lambda(q)$ as a
function of $q$ for introduced series. Filled square and circle
symbols show generalized Hurst exponent for river and corresponding
synchronized sunspot data set, respectively. Filled triangle symbol
indicates generalized scaling exponent, $\lambda(q)$. Weakly
$q$-dependency of scaling exponent implies the rivers and sunspot
numbers have almost multifractal behavior.  For all studied rivers,
the segments with fluctuations in the MF-DXA method near the mean
value are larger than that of far from the mean as shown in Figure
(\ref{dcca3}), namely the value of $\lambda(q)$ for $q<0$ is almost
larger than that of for $q>0$. This shows the statistics of small
fluctuations in water runoff are bigger than large fluctuations,
rare events. According to Figure (\ref{dcca3}), the value of
$\lambda(q)$ for $q>0$ remains almost constant, this indicates that
the rare events in the cross-correlation function behave in similar
way. On the other hand, the scaling behavior of fluctuation
function, $F_q(s)$ in the MF-DXA method confirms the various type of
fluctuations in different time scales have the same fractal
features. This may be useful to extend the fractal properties of
cross-correlation function which are derived in the short time
scales to larger one in the absence of large time series. In
addition, according to Figure (\ref{dcca3}), for $q=2$, we obtain
the empirical approximation, $\lambda(q=2)\approx [h_{{\rm
sun}}(q=2)+h_{{\rm river}}(q=2)]/2$.

According to auto-correlation function given by:
\begin{eqnarray}
C(\tau)&=&\langle [x(i+\tau)-\langle x\rangle][x(i)-\langle
x\rangle]\rangle \sim\tau^{-\gamma}
\end{eqnarray}
we can introduce the cross-correlation function for so-called long-range cross-correlation behavior as:
\begin{eqnarray}
C_{\times}(\tau)&=&\langle [x(i+\tau)-\langle x\rangle][y(i)-\langle
y\rangle]\rangle \sim\tau^{-\gamma_{\times}}
\end{eqnarray}
where $\gamma$ and $\gamma_{\times}$ are the auto-correlation and
cross-correlation exponents, respectively. Very often, direct
calculation of these exponents are not recommended due to the
non-stationarities and trends superimposed on the collected data.
One of the reliable and proper statistical methods to calculate
auto-correlation exponent is DFA method, namely $\gamma=2-2h(q=2)$
\cite{sadeghriver,physa}. Recently B. Podobnik et. al. have
demonstrated the relation between cross-correlation exponent,
$\gamma_{\times}$ and scaling exponent derived by equation
(\ref{Hq}) according to $\gamma_{\times}=2-2\lambda(q=2)$
\cite{DCCA}. The magnitude of cross-correlation and scaling exponent
derived by DFA and DCCA are reported in Table \ref{Tab1}.

\begin{figure} [t]
\epsfxsize=9cm\epsfbox{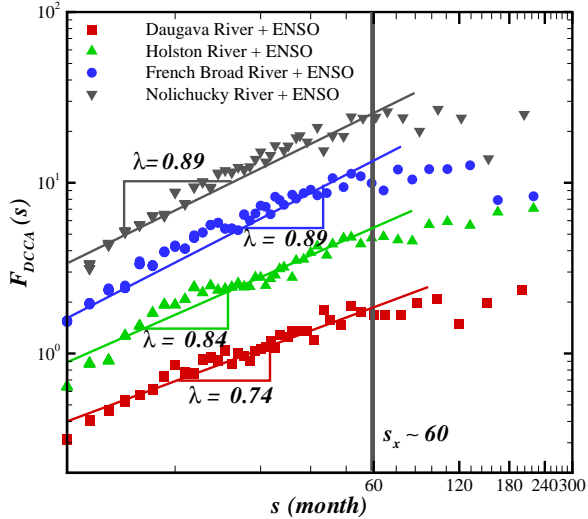} \narrowtext \caption{Fluctuation
function versus scale in log-log plot computed by DCCA method for
detrended river streamflows and El Ni\~{n}o index. To make more
sense we shifted the vertical scale.}
 \label{dcca_nino}
 \end{figure}

\begin{table}[htp]
\caption{\label{Tab1}Values of the scaling and cross-correlation
exponents of detrended sunspot numbers and river flow fluctuations using the
MF-DXA method for four rivers as well as scaling exponent given by the
DFA method, $h(q)$, of each synchronized data set for $q=2$.}
\begin{center}
\begin{tabular}{|c|c|c|c|c|}
  River&$h_{{\rm sunspot}}$&$h_{{\rm river}}$& $\lambda$& $\gamma_{\times}$ \\ \hline
       Nolichucky & $0.93\pm0.01$& $0.70\pm 0.01$&$0.85\pm0.01$& $0.30\pm 0.02$   \\\hline
          French Broad & $1.11\pm0.01$&$ 0.63\pm0.01$&$0.84\pm0.01$&$0.32\pm0.02$     \\\hline
     Holston & $1.05\pm0.01$& $0.60\pm0.01$&$0.78\pm0.01$&$0.44\pm0.02$     \\\hline
     Daugava & $1.14\pm0.01$& $0.61\pm0.01$&$0.76\pm0.01$&$0.48\pm0.02$     \\
\end{tabular}
\end{center}
\end{table}

To quantify the impact of El Ni\~{n}o (ENSO) phenomenon, we used the
El Ni\~{n}o $3$ index reported in \cite{enso} since 1950.  Applying
SVD method, the sinusoidal trends have been diminished, thereafter,
cleaned data sets used as inputs for DCCA algorithm. Figure
(\ref{dcca_nino}) indicates the results computed by DCCA method for
the detrended data sets. There is a crossover time scale in the
fluctuation function as a function of scale for all underlying
rives. The value of crossover equates to $s_{\times}\sim 60$ months
which is so-called the period of El Ni\~{n}o phenomenon. The value
of scaling exponents and cross-correlation exponents reported in
Table \ref{Tab2}, confirm that on $s\leq60$ months, there exists a
cross-correlation between ENSO phenomenon and rivers. For time scale
larger than $s>60$ months, this cross-correlation becomes ignorable
\cite{pablo,pablo10}. We also compare the impact of ENSO phenomenon
and sun activity on the streamflow of rives. To this end, we should
synchronize the El Ni\~{n}o, sunspot and river flow fluctuations.
The value of scaling exponents and cross-correlation exponents for
sunspot-river have also been reported in Table \ref{Tab2}. We find
that the contribution of sun activity represented by sunspot is
almost larger than ENSO phenomenon on the mentioned runoff rivers
during mentioned period.

\begin{table}[htp]
\caption{\label{Tab2}Values of the scaling and cross-correlation
exponents of synchronized El Ni\~{n}o 3 index and river flow
fluctuations as well as for sunspot and river in the same period,
using the DCCA method.}
\begin{center}
\begin{tabular}{|c|c|c|c|c|}
  River& $\lambda_{\rm ENSO}$& $\gamma_{\times}^{\rm ENSO}$ &$\lambda_{\rm sunspot}$&$\gamma_{\times}^{\rm sunspot}$ \\ \hline
       Nolichucky & $0.89\pm0.03$& $0.22\pm 0.06$ &$0.94\pm0.01$&$0.12\pm0.02$  \\\hline
          French Broad &$0.89\pm0.03$&$0.22\pm0.06$ &$0.98\pm0.02$&$0.04\pm0.02$    \\\hline
     Holston & $0.85\pm0.03$&$0.30\pm0.06$&$0.90\pm0.01$&$0.20\pm0.02$     \\\hline
     Daugava & $0.74\pm0.03$&$0.52\pm0.06$  &$0.77\pm0.01$&$0.46\pm0.02$   \\
\end{tabular}
\end{center}
\end{table}

\section{Discussion and conclusion}

There are many motivations such as complexity of fluctuations in our
environments which lead to examine them using robust and novel
methods from complex systems and statistical physics point of views.
Knowledge of natural variabilities are necessary to manage the
energy resources and prevent disasters from social and economical
point of views. In this paper, we analyzed three important
fluctuations in the nature, namely sun activity illustrated by
sunspot numbers, El Ni\~{n}o phenomenon and streamflow of rivers. However, many climate
indicators such as river flow fluctuations are affected by dominant
seasonal trends, but recent researches have confirmed that, there
are many variables causing the evolution of environmental conditions
behave as a complex systems. Subsequently, applying the common
methods in data analysis give incorrect or at least unreliable
results. On the other hand to infer valuable
results, the following necessary conditions should be satisfied:\\
i) The length of underlying time series should be large enough. \\
ii) The contribution of superimposed trends and noises on the
recorded data must be small enough or at least distinguishable.\\
Unfortunately above necessary conditions cannot be met in some
practical measurements. To solve these problems, we rely on the
robust methods in data analysis to explore the mutual effect of
sunspot and runoff water fluctuations. Based on the mentioned
motivation, we apply the most recent method, Multifractal Detrended
Cross-Correlation Analysis (MF-DXA), to examine the
cross-correlation and fractal properties of sunspot numbers and
river flow fluctuations for some most famous and available rivers.

Based on previous researches, as we expect, there are many
sinusoidal trends embedded in the underlying signals. Unfortunately,
these trends cannot be removed by common detrended procedures in DFA
and DCCA analysis \cite{kunhu,trend2}. These trends may cause some
spurious crossovers in the fluctuations function versus time scale
related to the competition between noise and trends. According to
Figure (\ref{crossover}), there are some crossovers in the
fluctuation function of the sunspot numbers versus time scale,$s$.
The first one corresponds to the annually period, the second is
equal to the so-called solar activity, 11 years, while the last one
may indicate the well-known Gleissberg period, but we must point out
that this value with small uncertainty cannot be determined by this
current data, because the size of data set is small. Here to
eliminate those trends, we use the Singular Value Decomposition
(SVD) method.

Results given by DFA method for original runoff rivers and sunspot
numbers, demonstrated that all underlying runoff river behave as the
anti-persistent long-range correlated series for time scale $s<12$
months at $1\sigma$ confidence interval (Figure (\ref{dfa})). On the
other hand, the value of Hurst exponents for series without
sinusoidal trends exhibit runoff rivers and sunspot numbers are
long-range correlated process (see Table \ref{Tab1}).

MF-DXA analysis for non-detrended underlying data sets implied
universal scaling exponent of fluctuation function for $q=2$ for all
underlying rivers flow at intermediate time scales, $[12-24]\leq s
\leq 130$ months, namely $\lambda=1.17\pm0.04$ (see Figure
(\ref{fs_q})). This finding has recently been reported in Ref.
\cite{Zanchettin08} from power spectrum analysis. On the other hand
there is a similar crossover in cross-correlation fluctuation
functions at $s_{\times}\sim 130$ months which is known as cycle of
solar activity, for all studied rivers versus sunspot
\cite{sadeghriver}.

Data without sinusoidal trends were used as the inputs for DCCA and
its generalized MF-DXA methods. Due to the scaling behavior of
fluctuations function (equation (\ref{Hq})) versus time scale, $s$,
we concluded that there exists a significant cross-correlation
between sun activity and water runoff river fluctuations. This
cross-correlation confirms that the influence of other reasons on
the river flow fluctuations such as human activity, drainage network
morphology, land use patterns and topography may be ignored or at
least cannot be distinguished form major effect which here is sun
activity \cite{Zanchettin08}. It must point out that, we know that
the amount of radiation given off by the Sun (solar irradiance) is
greatest when there are lots of sunspots. As discussed in detail in
Ref. \cite{pablo}, the higher value of galactic cosmic ray, the
higher the cloud cover which increases water resource of river, so
we expect streamflow to be almost cross-correlated with solar
activity. Other explanation are also investigated in
\cite{bha07,Ruz06}.

The cross-correlation exponent, $\gamma_{\times}$, of underlying
data set in the presence of non-stationarities and trends have been
determined by DCCA method. According to the relation between scaling
exponent, $\lambda(q=2)$, and cross-correlation exponent,
$\gamma_{\times}$,  namely $\gamma_{\times}=2-2\lambda(q=2)$, one
find that for long-range cross-correlated signals, the value of
$\gamma_{\times}$ goes to small value demonstrating the
cross-correlation function decreases more slowly and statistically
two underlying data sets tend their present situation in their
cross-correlation behavior. According to the values reported in
Table \ref{Tab1} and Figures (\ref{dcca1}) and (\ref{dcca2}), there
exists a long-range cross-correlation between rivers flow
fluctuations used in this paper and sun activity indicated by
sunspot numbers.

Weak $q$-dependency of the generalized Hurst, $h(q)$, and scaling,
$\lambda(q)$, exponents, demonstrates that the original underlying
series and also cross-correlation between sunspot numbers and each
river behave as almost multifractal processes, demonstrating another
universal characteristics. The value of $\lambda(q)$ for $q>0$
remains almost constant, this indicates that the rare events in the
cross-correlation function behave in similar way. The similarity in
behavior of fluctuation function, $F_q(s)$ in the MF-DXA method
demonstrates the various type of fluctuations in different time
scales have the same fractal features. Based on this result, this
may be useful to extend the fractal properties of cross-correlation
function derived by using short time scales to larger one without
relying on the large time series. In addition, the rare events in
streamflow and sunspot cross-correlation analysis, are not affected
by local characteristics.

 According to
our results, the empirical relation between $\lambda(q=2)$ and Hurst
exponent, $\lambda(q=2)\approx [h_{{\rm sun}}(q=2)+h_{{\rm
river}}(q=2)]/2$, has also been confirmed.

Since the value of cross-correlation of Daugava river and sunspot
numbers is less than other rivers, one can conclude that, beside the
sun as a main resource of energy in the nature, the river's
geographical situation and the source of rainfall for that river may
have reasonable impact on runoff water fluctuation. Recently, even
an opposite behavior between  solar activity and rainfall
fluctuation in equatorial east Africa  has been noticed in the
recent report of the East Africa \cite{africa} explained in Ref.
\cite{pablo}. In addition, the value of $\lambda$ for French
Broad and Nolichucky rivers are larger than two remained runoff
rivers. Figure (\ref{power_all}) also indicated, the
power spectrum of French Broad and Nolichucky rivers behave in
similar way around the well-known solar activity period.  We also found a sharp
crossover in the fluctuation functions derived by MF-DXA method for
mentioned rivers in Figure (\ref{all_q}).  Since, the geographical positions of French Broad and
Nolichucky rivers are close together, subsequently as expressed
before, the impact of sun activity represented by sunspot, depends
on the geographical properties of rivers.

One of the most important phenomenon which can affect the runoff
rivers is El Ni\~{n}o (ENSO). To compare the contribution of ENSO
 phenomenon and sun activity, we used El Ni\~{n}o index $3$. The results derived
by DCCA have been indicated in Figure (\ref{dcca_nino}) and also
reported in Table \ref{Tab2}. There exists a crossover time scale in
the log-log plot of fluctuation function versus scale. The value of
this crossover time scale is $s_{\times}\sim60$ months regarding to
well-known ENSO period. ENSO phenomenon is cross-correlated by
streamflow fluctuations for $s\leq60$ months, while for larger than
this period, one can ignore this cross-correlation behavior. By
comparison, the effect of ENSO phenomenon and sun activity on the
river flows, we found that the contribution of sunspot is almost
larger than ENSO index since $1950$.

There are many methods to predict the sun activity \cite{pc,li}, so
based on our current results which is demonstrated the
cross-correlation between detrended sunspots and water runoff, one
may use those results to predict river flow fluctuations.

Finally, one must point out that it should be interesting to extend
the present analysis to a various set of runoff water fluctuations
and other sun activity indicators such as solar irradiance, galactic
cosmic rays flux and geomagnetic activity to find whether the values
for the main parameters used in this analysis have a more universal
validity of the solar influence on the streamflow fluctuations
assigned as a climate variable in this paper.

{\bf Acknowledgements} This research has been financially supported
by Shahid Beheshti University research deputy affairs under grant
No. 600/405.


\end{document}